\documentclass[pre,reprint,floatfix,amsmath,amssymb,twocolumns]{revtex4-1}
\usepackage{float}
\usepackage{amssymb, amsmath}
\usepackage{booktabs}
\usepackage[colorlinks,citecolor=blue,urlcolor=blue]{hyperref}
\usepackage{color}
\usepackage{graphicx}
	\newcommand{\InsertFig}[4]
	{\begin{figure}[h!t]
			\centerline{
				\includegraphics[width=#4\columnwidth]{./#1}
			}
			\caption{{\footnotesize  #2}
				\label{fig:#3}}
		\end{figure}}
	\newcommand{\InsertFigTwo}[5] {
		\begin{figure*}[h!t]
			(a)
			\includegraphics[width=#5\textwidth]{./#1}
			(b)
			\includegraphics[width=#5\textwidth]{./#2}
			\caption{{\footnotesize  #3}
				\label{fig:#4}}
		\end{figure*}}	
\newcommand{\InsertFigFour}[7] {
				\begin{figure*}[h!t]
					(a)
					\includegraphics[width=#7\textwidth]{./#1}
					(b)
					\includegraphics[width=#7\textwidth]{./#2}
					
					(c)
					\includegraphics[width=#7\textwidth]{./#3}
					(d)
					\includegraphics[width=#7\textwidth]{./#4}
					\caption{{\footnotesize  #5}
						\label{fig:#6}}
				\end{figure*}}

\def\Eq#1{Eq.$\,$(\ref{eq:#1})}

\def\ts{\textstyle}
\def\half{{\ts{1\over2}}}

\newcommand{\Fig}[1]{Fig.~\ref{fig:#1}}

\begin{document}

\title{Quantum correlations for a simple kicked system with mixed phase space}
\author{Or Alus}
\email{oralus@tx.technion.ac.il}
\affiliation{Department of Physics,Technion- Israel Institute of Technonlogy, Haifa, Israel 3200003}

\author{Shmuel Fishman}
\email{fishman@physics.technion.ac.il}
\affiliation{Department of Physics,Technion- Israel Institute of Technonlogy, Haifa, Israel 3200003}

\author{Mark Srednicki}
\email{mark@physics.ucsb.edu}

\affiliation{Department of Physics, University of California, Santa Barbara, CA 93106}

\begin{abstract}
We investigate both the classical and quantum dynamics for a simple kicked system (the standard map) that classically has mixed phase space. For initial conditions in a portion of the chaotic region that is close enough to the regular region, the phenomenon of sticking leads to a power-law decay with time of the classical correlation function of a simple observable. Quantum mechanically, we find the same behavior, but with a smaller exponent. We consider various possible explanations of this phenomenon, and settle on a modification of the Meiss--Ott Markov tree model that takes into account quantum limitations on the flux through a turnstile between regions corresponding to states on the tree. Further work is needed to better understand the quantum behavior. 
\end{abstract}

\maketitle
\section{Introduction}
The subject of this work is the long term correlations for mixed dynamical systems. For such systems,
the motion is chaotic in some regions of the classical phase space, while in other regions it is regular \cite{lichtenberg1983,Tabor1989,zaslavsky2005hamiltonian}. 
An important phenomenon such systems exhibit is sticking:
trajectories in the chaotic region that are close to the perimeter of a regular region will stay close for a long time.
This can be quantified by the survival probability, the probability for a trajectory to be near its initial point after
a time $t$; sticking results in a a power-law decay of the survival probability with time.
Systems with mixed dynamics
include the H\'enon map \cite{Henon1969}, the standard map \cite{Zumofen1994,Chirikov1979a}, and the ``molecular cat'' \cite{Katz2017}. The classical behavior in such situations was studied extensively in the past \cite{Channon1980,Chirikov1984,Meiss1985,Karney1982,Karney1983,MacKay1984,Zaslavsky1997,Ceder2013}.

In the present work we focus on the quantum mechanical behavior, and emphasize how it differs from the classical.
We study the standard map on a torus, with parameters that would yield transport dominated by accelerator modes
on a cylinder; these
are phase space islands for which momentum grows linearly in time \cite{Zumofen1994,Zaslavsky1997,Ishizaki1991,Chirikov1979a,Manos2014}.
The dynamics around these islands is modeled by the H\'enon map \cite{Karney1982,Karney1983}. 

In earlier work some of us explored the survival probability in the framework of the Meiss--Ott Markov tree model. In particular it was found that the model predicts correctly the decay of the survival probability, falling of in time with a power 
\begin{equation}\label{eq:powerlaw}
P_{Sur}(t)\sim \frac{1}{t^\gamma}
\end{equation}
where $\gamma$ does not depend systematically on the parameters of the standard map or the related H\'enon map. It takes the value 
\begin{equation}\label{eq:exp}
\gamma\approx 1.57\;,
\end{equation}
as was found with various methods \cite{Alus2017,Alus2014,Cristadoro2008,Frahm2013}. 
On the torus,
the accelerator-mode islands reduce to stationary islands, since the motion is periodic in both position and momentum directions.

In the present work we study the two-point time correlation function of a simple observable. Such a correlation function
has a straightforward definition both classically and quantum mechanically, which facilitates comparison of their behavior.
Classically, for initial conditions localized in the sticking region of the chaotic component, the correlation function
will oscillate without decay, and then have
a power-law decay that is related to the power-law decay of the survival probability. 
Quantum mechanically, we also find a power-law decay, but one that is slower for sufficiently large values of 
Planck's constant.
We examine several possible mechanisms for this quantum slowing down, and conclude that quantum effects lead to ``pruning'' and ``truncation'' of the Meiss--Ott Markov tree, resulting in strong deviation from the power law in the decay of the survival probability and of the correlation function. 

The outline of the paper is as follows. In Sec. \ref{sec:Decay}  the quantum and classical correlation functions are defined and computed, and their behavior compared.
In Sec. \ref{sec:spread} we consider some specific mechanisms of quantum trapping, and conclude these are irrelevant for the present case. We propose a mechanism for modification of the Markov tree by quantum effects and conjecture the resulting form of the decay of the correlations function. Our conclusions are in Sec. \ref{sec:Conc}.

\section{Decay of Quantum and classical correlation functions}\label{sec:Decay}
The standard map is \cite{Chirikov1979a,lichtenberg1983}
\begin{align}
p_{t+1} &= p_t+K\sin(x_t)
\label{map1} \\
x_{t+1} &= x_t+p_{t+1}\;.
\label{map2}
\end{align}
We take both $x$ and $p$ to be periodic with period $2\pi$ so it is defined on a torus. The Hamiltonian that generates this map is
\begin{equation}
H = {\ts{1\over2}} p^2 + K\cos(x){\ts\sum_n}\delta(t-n) \;.
\label{h}
\end{equation}
To quantize the map while preserving both periodicities, we discretize both $x$ and $p$, as is done for the 
baker's map or cat map \cite{Balazs1986,Balazs1989,Hannay1980}. Mapping
\begin{align}
x &\to x_j = 2\pi (j+\eta)/N\,,\quad j=0,\ldots,N-1\;,
\label{d1} \\
p &\to p_k = 2\pi (k+\eta)/N\,,\quad k=0,\ldots,N-1\;,
\label{d2}
\end{align}
where $0\le\eta<1$ is a possible offset, and $N$ is a positive integer. 
We should have $\exp(ipx/\hbar)=\exp(2\pi i (j+\eta)(k+\eta)/N)$, which implies
\begin{equation}
\hbar = {2\pi\over N}\;.
\label{hbar}
\end{equation}
The inner product between a position eigenstate and a momentum eigenstate is then
\begin{equation}
\langle x_j|p_k\rangle ={1\over\sqrt{N}}\exp(2\pi i(j+\eta)(k+\eta)/N)\;.
\label{<x|p>}
\end{equation}
Position and momentum eigenstates each form a complete basis,
\begin{equation}
\sum_{j=0}^{N-1}|x_j\rangle\langle x_j|
=\sum_{k=0}^{N-1}|p_k\rangle\langle p_k|
=I\;.
\label{<x|p>}
\end{equation}
The unit-time evolution operator is
\begin{equation}
U = \exp(-i p^2/2\hbar)\exp(-i K \cos(x)/\hbar)\;.
\label{U}
\end{equation}
Using eqs.$\,$(\ref{d2}) and (\ref{hbar}), $\exp(-ip^2/2\hbar)\to\exp(-i \pi (k+\eta)^2/N)$.
We want this to be invariant under under $k\to k+N$.
This requires $\eta+\half N$ to be an integer. So we must have either $\eta=0$ and $N$ even,
or $\eta=\half$ and $N$ odd. For simplicity with the numerics, we choose
\begin{equation}
\eta=0\;,\quad N\ \hbox{even}\;.
\label{eN}
\end{equation}
We also choose to let the $2\pi$-periodic coordinates $x$ and $p$ be in the range $[-\pi,\pi]$ rather than $[0,2\pi]$ 
so that the classical island is in the center of the range.

We choose $\cos(x)$ as the observable whose correlations we will study. We have chosen this function since its phase space average vanishes.
We wish to calculate the quantum correlation function, and compare it to the classical result. 
For this purpose we define
\begin{align}
\tilde{f}_Q(t) &=\langle \cos({x_t})\cos({x_0})\rangle\cr
&=\langle\Psi_0| (U^\dagger)^t \cos(x)(U)^t \cos(x)|\Psi_0\rangle\;,
\end{align} 
where ${x_t}$ is the position operator in the Heisenberg representation at time $t$.
For the convenience of calculation, we split the computation into two parts 
\begin{equation}
|R\rangle=(U)^t \cos({x})|\Psi_0\rangle
\end{equation}
and
\begin{equation}
\langle L| = \langle\Psi_0| (U^\dagger)^t \;.
\end{equation}
Here  $\langle L|$ and $|R\rangle$ are vectors in the Dirac representation.
These expressions are calculated using FFT starting from  $ \langle\Psi_0|$ and $|\Psi_0\rangle$ . 
The product is computed to obtain 
\[
\tilde{f}_Q(t)= \langle L| \cos({x})|R\rangle\;.
\]
To obtain a real correlation function that can be compared to its classical counterpart, we define the symmetrized
product
\[
\tilde{f}_S(t)= \frac{1}{2}\left[ \cos({x_t})\cos({x_0})+ \cos({x_0})\cos({x_t})\right]\;.
\]
Its expectation value is 
\begin{equation}
f_Q(t)=\langle\Psi_0|\tilde{f}_S(t)|\Psi_0\rangle=\Re\left[\tilde{f}_Q(t)\right]\;.
\end{equation}
In the classical limit, it is expected to approach 
\begin{equation}\label{eq:ClassicCorr}
f_C(t)=\langle\cos(x_t)\cos(x_0)\rangle_C\;.
\end{equation}
where $x_t$ is the classical position at time $t$ and $\langle\rangle_C$ is the average over the classical initial conditions corresponding to $|\Psi_0\rangle$. 

We consider values of $N$ from $2^{17}$ to $2^{21}$, up to time $t=10^4$. 
As the initial condition we take a minimum-uncertainty quantum wave packet,  
centered on a phase-space point $(x_c,p_c)$, which need not be a grid point. In the position representation it takes the form  
\begin{align}\label{eq:wavepacket}
	\psi(x_j) &\propto \exp[  i p_c x_j/\hbar -(x_j-x_c)^2/2\hbar]\cr
	&\propto \exp[ i N p_c x_j/2\pi -N(x_j-x_c)^2/4\pi]\cr
	&\propto \exp[ i p_c (j+\eta) -\pi N(j+\eta-x_c/2\pi)^2]\,.
\end{align}
We take $x_c=1.7 ,\; p_c=-0.25$. 
The corresponding classical phase-space density is
\begin{align}\label{eq:Classicalpacket}
	\rho(x,p) &\propto \exp[-(x-x_c)^2/\hbar-(p-p_c)^2/\hbar]\cr
	&\propto \exp[-N(x-x_c)^2/2\pi-N(p-p_c)^2/2\pi]
\end{align}
For parameter value of $K=6.476939$ this wave packet is centered around a hyperbolic fixed point of a secondary island chain belonging to the ``accelerator island''. As explained in the introduction, the island corresponds to an accelerator mode if the map is defined on a cylinder, but in the present paper it is defined on a torus and the island is stationary. The classical ensemble consists of $N_0=10^6$ initial conditions. The initial point was chosen so that it is in a favorable situation to observe the power law decay of correlations resulting of sticking. The islands belonging to accelerator modes are approximated by the H\'enon map \cite{Karney1982}. We prepared the initial distribution centered on the hyperbolic point of a period 5; see Fig.~\ref{fig:fig1}. The initial wave packet (and the corresponding distribution of the initial condition) is much narrower than the size of the torus. At the initial stages (for $t<10^2$)  we see rapid oscillations of period 5. At later time only sticking trajectories stay at the vicinity of the island. We assume following Karney \cite{Karney1983} that only these contribute to the long time behavior of the correlation function $f_C(t)$, while the rest diffuse in phase space and do not contribute to the correlation function. We assume that the probability to stick for a time $\tau$ is for long $\tau$
\begin{equation}\label{eq:Survivaldens}
p(\tau)\sim \frac{1}{\tau^{\gamma+1}}
\end{equation}
The survival probability is asymptotically 
\begin{equation}\label{eq:Survival}
P_{Sur}(t)\sim\int\limits_{t}^{\infty} \frac{1}{\tau^{\gamma+1}} d\tau \sim \frac{1}{t^\gamma}\;,
\end{equation}
with the value of $\gamma$ given by \Eq{exp}.
We turn now to the calculation of  the decay of the classical correlation function \Eq{ClassicCorr}. Using the assumption that only the sticking part contributes to the correlation function
\begin{equation}\label{eq:Correlations}
f_C(t)\sim \int\limits_{t}^{\infty}\tau \frac{1}{\tau^{\gamma+1}} d\tau \sim   \frac{1}{t^{\gamma-1}} \;.
\end{equation}
We used the fact that the probability that $0$ and $t$ are both in an interval of length $\tau$ is proportional to $\tau$. 
We see that the decay of the classical correlation function is strongly related to the decay of the survival probability. 

In order to compare with the quantum sticking we prepare an initial wave function corresponding to a minimal uncertainty wave packet (\ref{eq:wavepacket}). In \Fig{AlgCorr} We compare the correlation function with the one found for the corresponding classical density  (\ref{eq:Classicalpacket}). We find a region of oscillations with period 5  for $t<5\cdot10^2$, while for $5 \cdot 10^2<t<5\cdot 10^3$ 
we see a regime of power-law decay with exponent $\gamma-1\approx 0.8$, in reasonable agreement with \Eq{exp}. We did not calculate the correlation function beyond $10^4$ since up to that value we found excellent agreement between the quantum and classical results for $N=2^{21}$, as is clear from \Fig{AlgCorr}. For longer times  the quality of the classical distribution with $N_0=10^6$ initial conditions deteriorates, since only a small fraction of the trajectories does not run away from the sticky region. Running for much larger ensemble is beyond our numerical possibilities. Moreover we believe that the current results are sufficient for the conclusions of the paper.

In Fig.~\ref{fig:hbar} we present the quantum correlation function for various values of Planck's constant $\hbar=\frac{2\pi}{N}$. We find that as $\hbar$ increases the quantum correlation function decreases slower than the classical one. To understand it we plot in \Fig{ClassicCorrFig} the classical correlation as found starting from initial conditions corresponding to the ones used for the correlation function presented in \Fig{hbar}. The initial conditions are patches of the blue heavy points near the hyperbolic fixed point in \Fig{fig1}.
If the initial condition was a distribution satisfying detailed balance, the classical survival probability would decay with the exponent $\gamma-1$ \cite{Meiss1997}. For our initial distribution there is no definite theoretical prediction, but since it is within the strictly chaotic layer we expect it to decay like $\gamma -1$ approximately. Indeed we find this exponent to vary in the interval $\gamma-1\in[0.5,0.75].$ 
We consider this numerical result to be in agreement with theory.
The quantum correlations as can be seen from \Fig{hbar} decay more slowly, with exponents $\gamma_q-1=0.1,0.15,0.33,0.49,0.81$ for 
$N= 2^{17},2^{18},2^{19},2^{20},2^{21}$ respectively. The fits were performed in the interval $t\in [5\cdot 10^2,5\cdot 10^3]$ using log-log scale with uniform distribution of points. The rest of this paper is devoted to the explanation of the difference between the quantum and the classical distributions. 

\InsertFigTwo{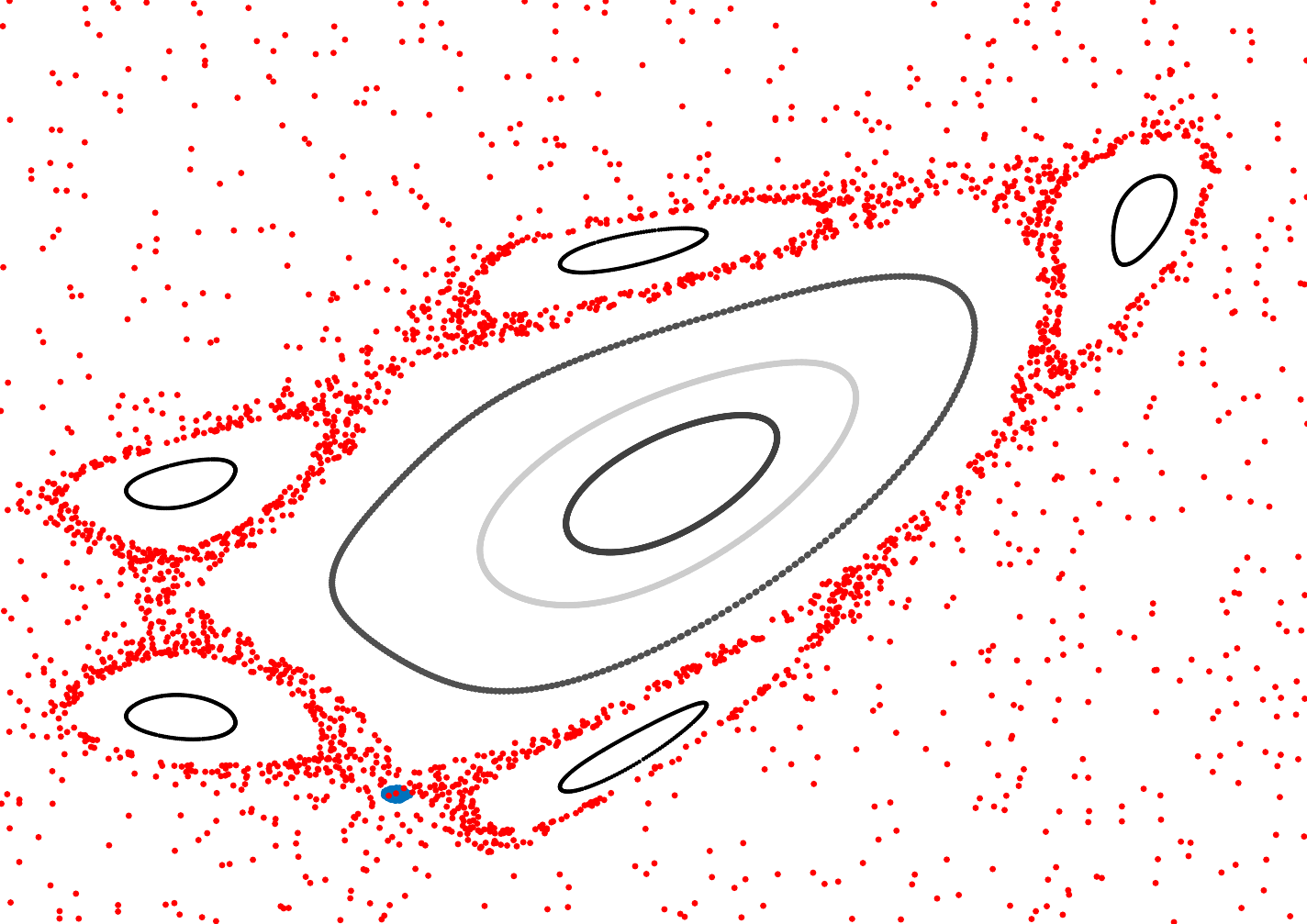}{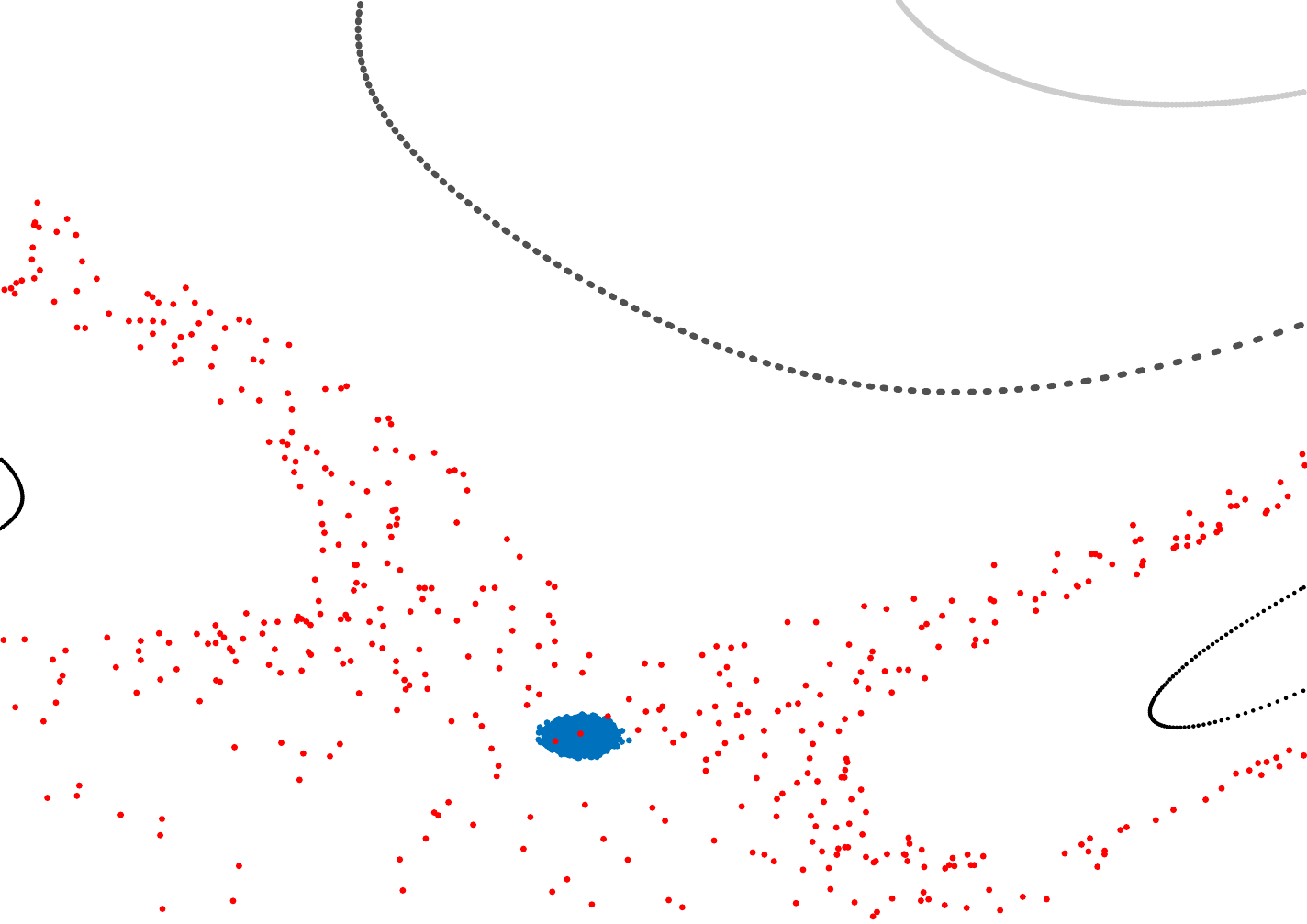}{(a) The island for $K=6.476939$ and the initial classical phase space density (in heavy blue) corresponding to $N=2^{21}$.  (b) The same as (a) but zoomed around the initial density.}{fig1}{0.4}
\InsertFig{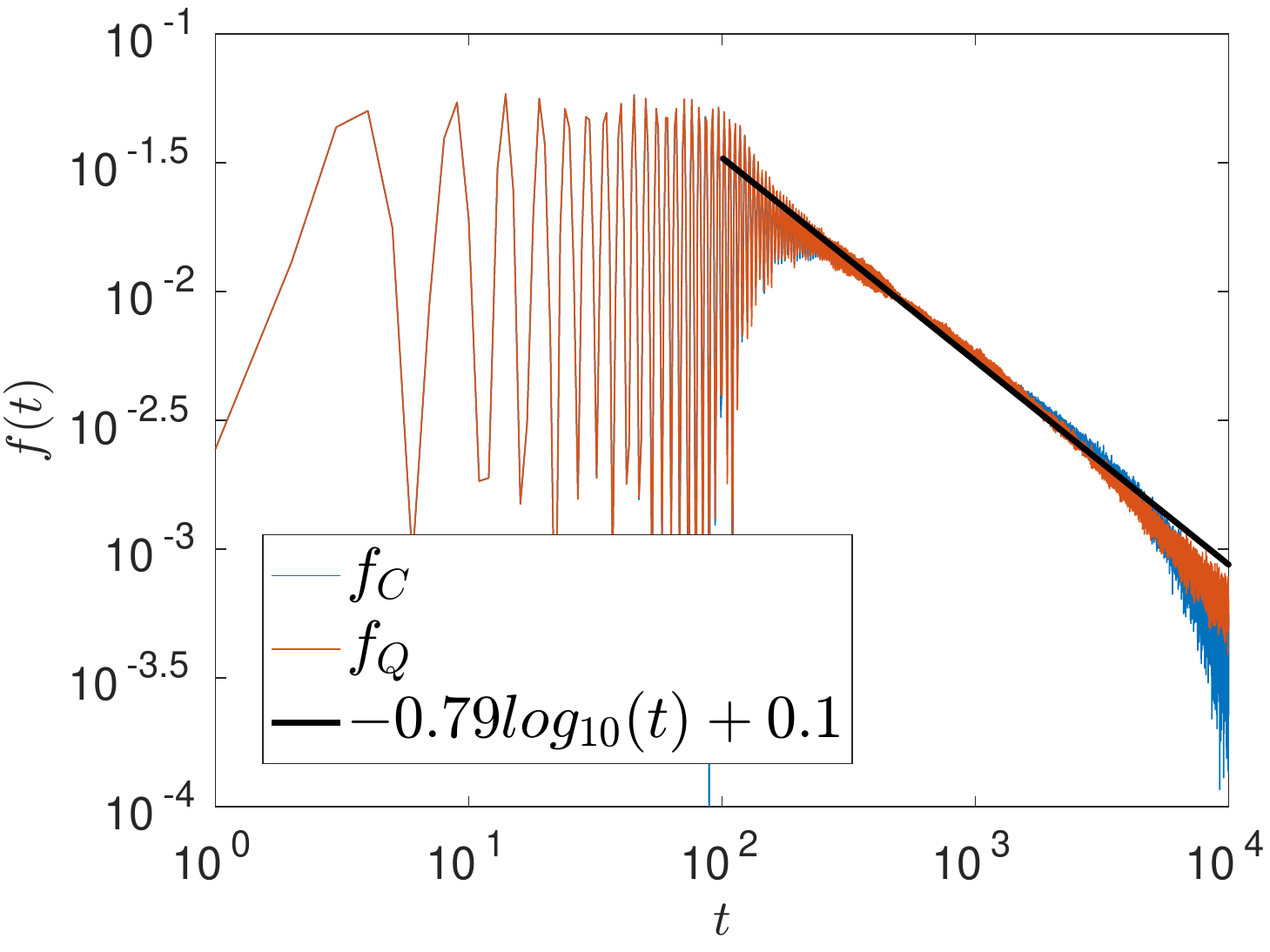}{The classical and quantum correlation functions for $K=6.476939$, $N=2^{21}$, and 
	the initial conditions of \Fig{fig1} on a log-log plot. For comparison the straight line with $y=-(\gamma-1)x+b$ where $\gamma-1$ is the exponent of the power law decay resulting from the best fit for$5\cdot10^2<t<5\cdot10^3$.}{AlgCorr}{0.8}
\InsertFig{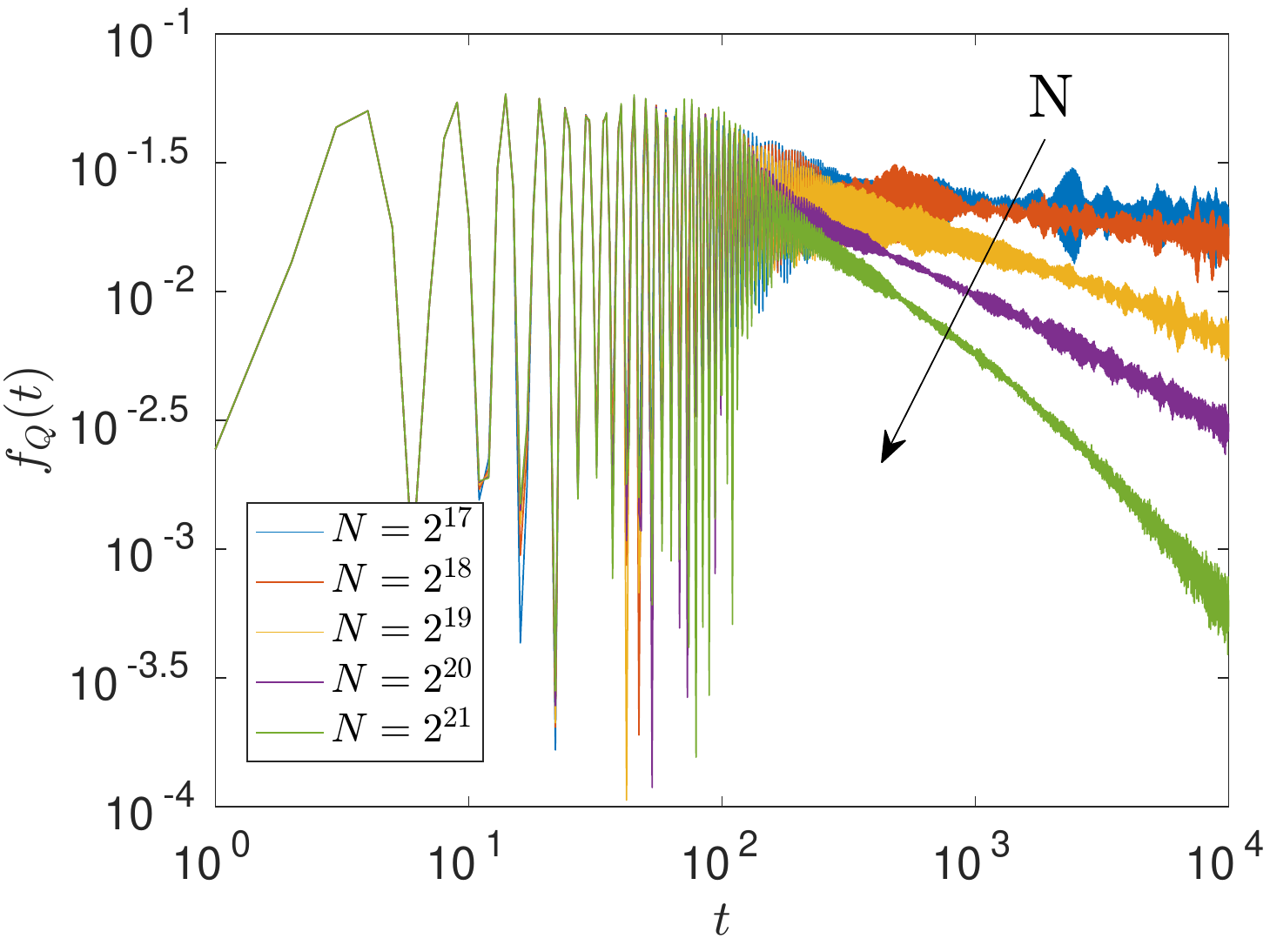}{The quantum correlation function for the initial conditions of \Fig{AlgCorr}, but for 
	$N=2^{17}{-}2^{21}$.}{hbar}{0.8}
\InsertFig{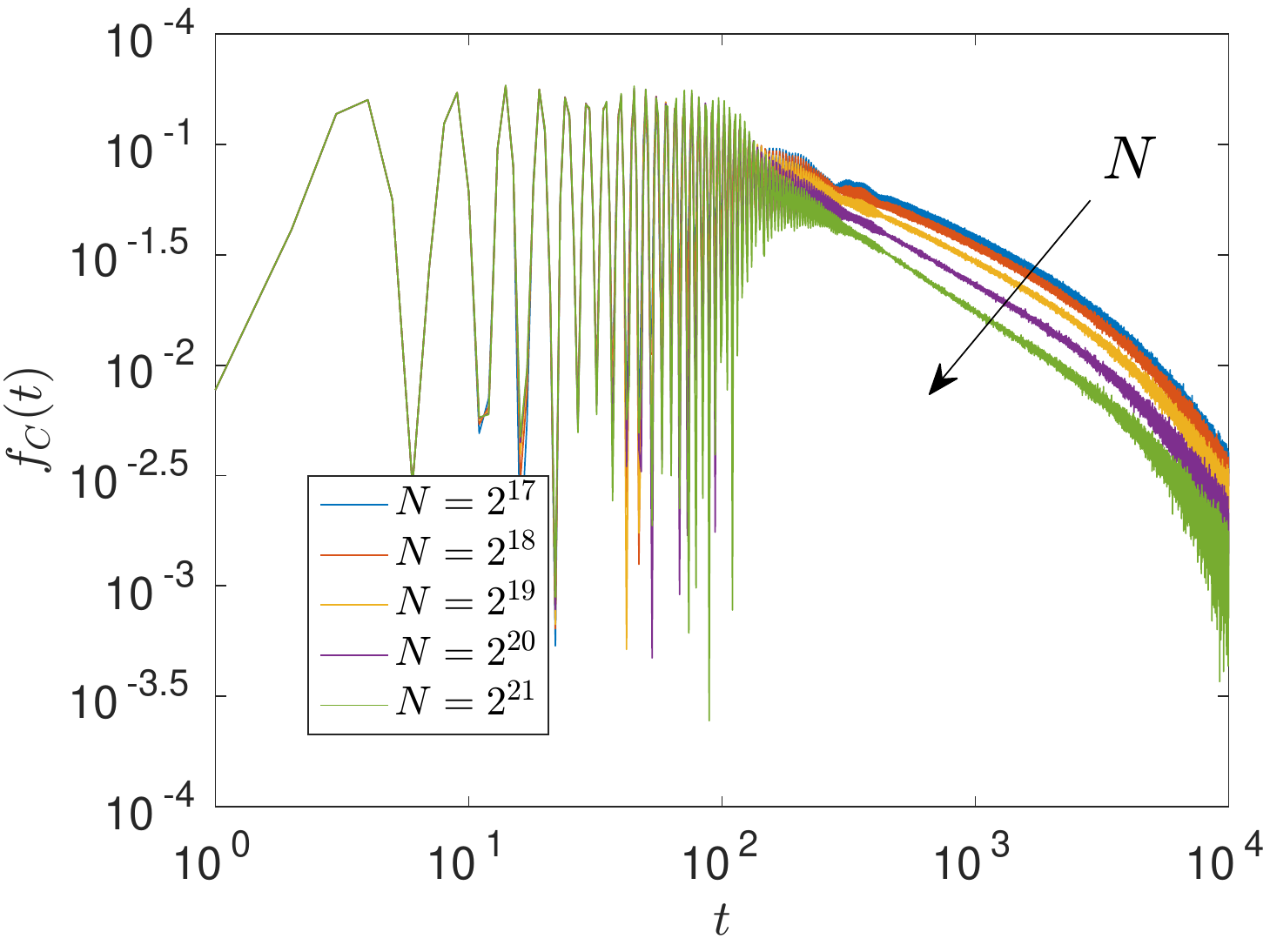}{The classical correlation functions corresponding to the quantum ones presented in \Fig{hbar} The dependence on $N$ results from different initial conditions chosen to match the quantum initial wave function}{ClassicCorrFig}{0.8}

\section{Detailed analysis of the quantum spreading}\label{sec:spread}
We would like to understand what  slows the quantum spreading compared to the classical one. For this purpose we analyze the mechanisms that slow down quantum spreading. The first is trapping in islands. For this purpose we use the Husimi function defined as 
\begin{equation}\label{eq:Husimi}
\mathcal{H}(x',p') = \frac{1}{2\pi \hbar}\left|\intop_{-\pi}^{\pi} dx\, \psi(x) \frac{1}{\sqrt[4]{\pi \hbar}}e^{-\frac{(x-x')^2}{2\hbar}+\frac{ip'x}{\hbar}}\right|^2\;,
\end{equation} 
which is normalized such that \[
\intop_{-\pi}^{\pi}\intop_{-\pi}^{\pi} \mathcal{H}(x',p')dx' dp'=1\;.
\]

Note that the wave function has finite weight inside the classical island (see Fig. \ref{fig:island}). We fix the grid of $p',x'\in[-\pi,\pi]$ to be $1024\times1024$. This is done mainly due to computation time and memory considerations. Note that the grid of $x$ from \Eq{Husimi} is still set by $\hbar$. 

In order to understand the different power laws, we define a weight $W_\mathcal{D}$ as the integral of the Husimi distribution over the domain $\mathcal{D}$, 
\begin{equation}\label{eq:Weq}
W_\mathcal{D} = \intop_\mathcal{D}\intop   dp' dx' \mathcal{H}(x',p')\;,
\end{equation}
where several different domains $\mathcal{D}$ are chosen in what follows. 
We study the behavior of $W_\mathcal{D}$ for different values of $N$ (and therefore $\hbar$). The normalization of the Husimi function is such that in the classical limit the weights reduce to the survival probability of Eq.(1) of \cite{Alus2017}.
\subsection{Trapping in islands} 
In order to study the effect of trapping inside the classical islands we study the Husimi distribution in a classical island. This is shown in \Fig{island}  for time $t=10^4$.
The most external classical trajectory (marked by a heavy black line) is the boundary circle of the island. It is determined by following classical trajectories started inside the island. Using the island as a domain of integration for the Husimi function we calculate by \Eq{Weq} the weight in small and big islands $W_s$ and $W_b$ respectively. These are plotted in \Fig{weightsHyp}. We see that these weights saturate at $t=10^2$ and stay constant at least till $t=10^4$, and correlations presented in \Fig{hbar} decay for shorter time, therefore this process is irrelevant for the understanding of the decay of correlations presented in \Fig{hbar}. 

We checked also that if the initial wave packet is placed in the small island the weight in the small and in the big island does not decay significantly at least till $t=10^4$, therefore it is irrelevant for the result of \Fig{hbar}.  
Similar behavior was seen in \cite{bittrich2014temporal,Backer2005}. In our case the build up in the island is faster than in the cited paper, since our initial wave packet is in the sticky region, near the hyperbolic fixed point. 

\InsertFigFour{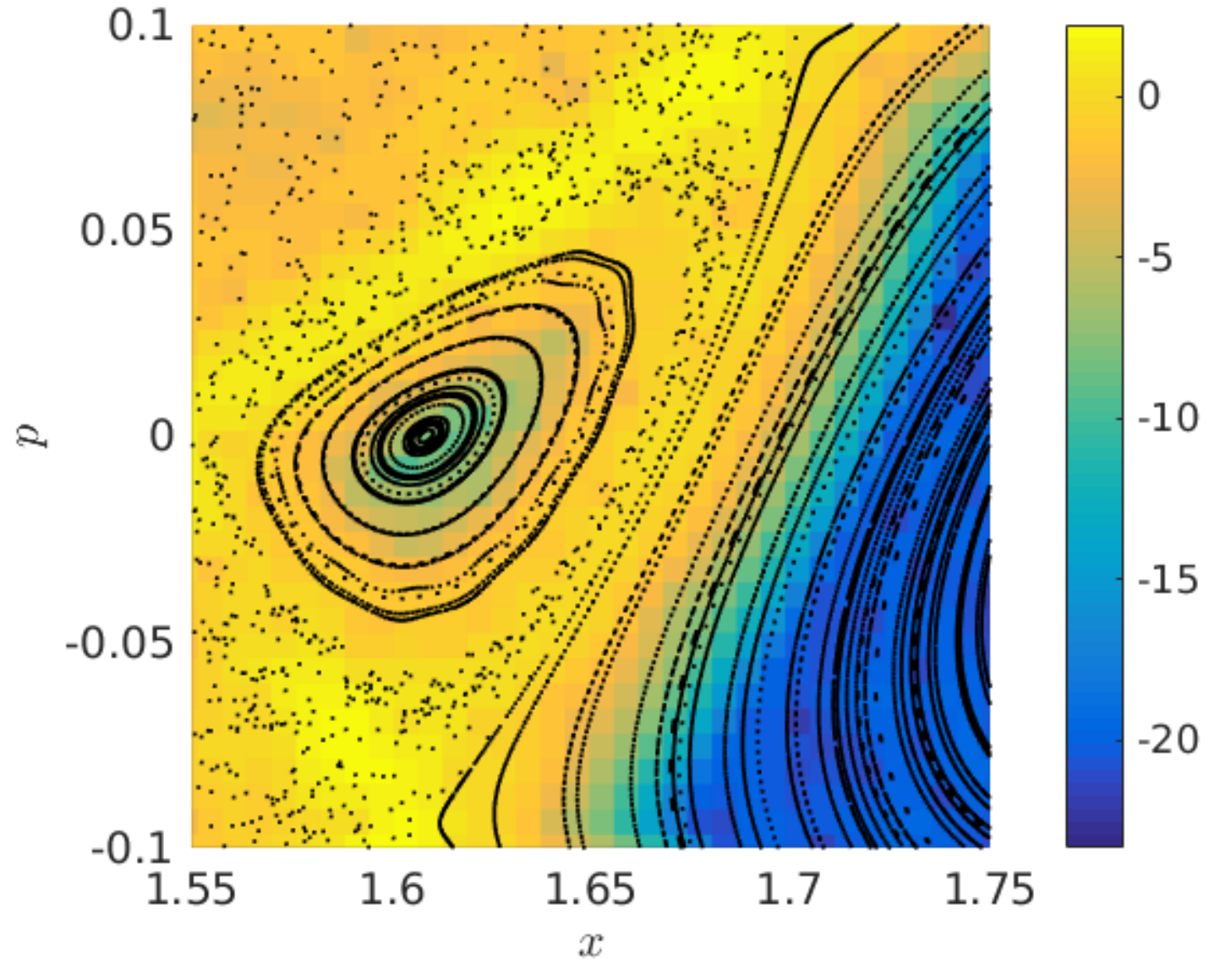}{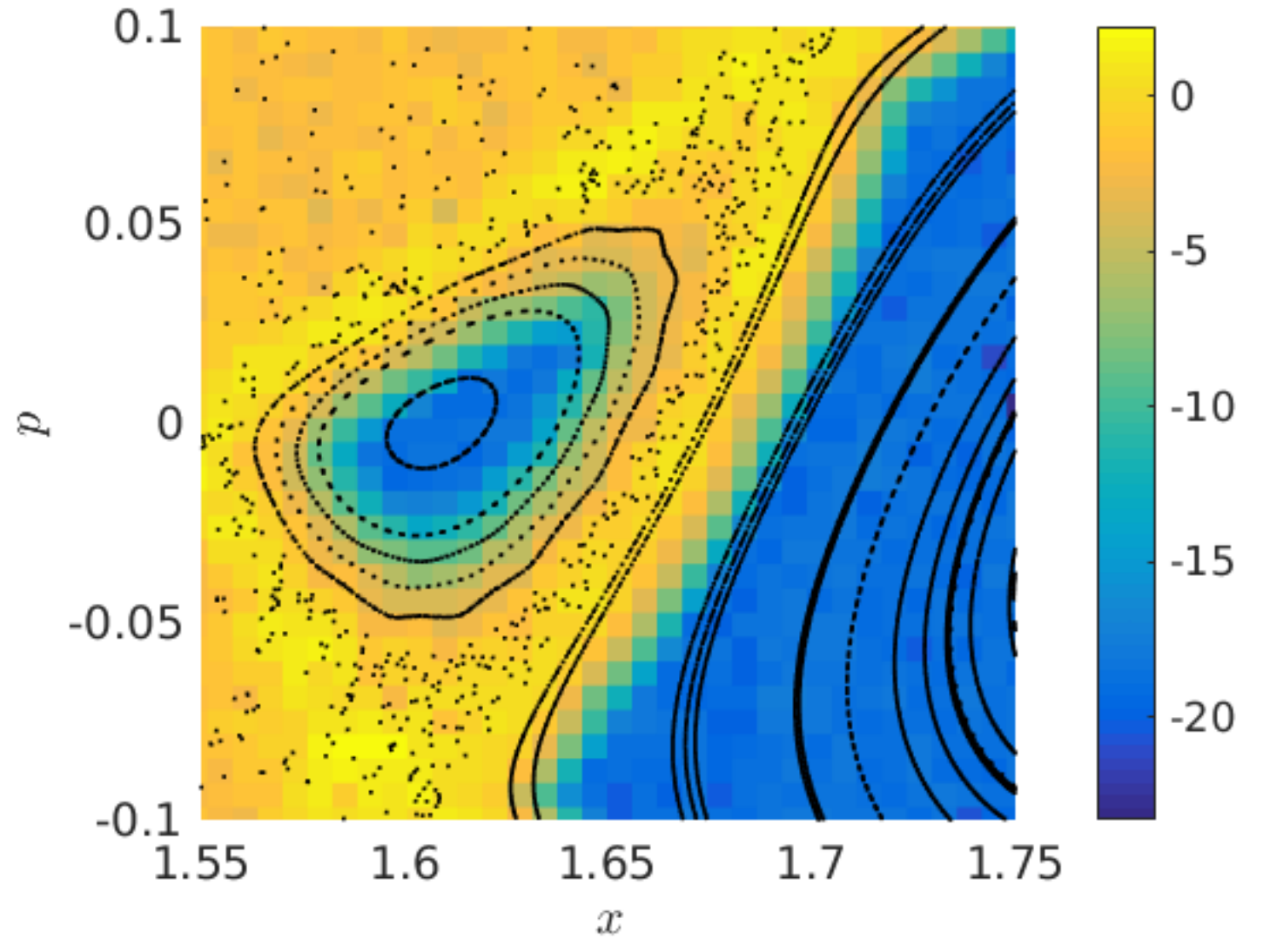}{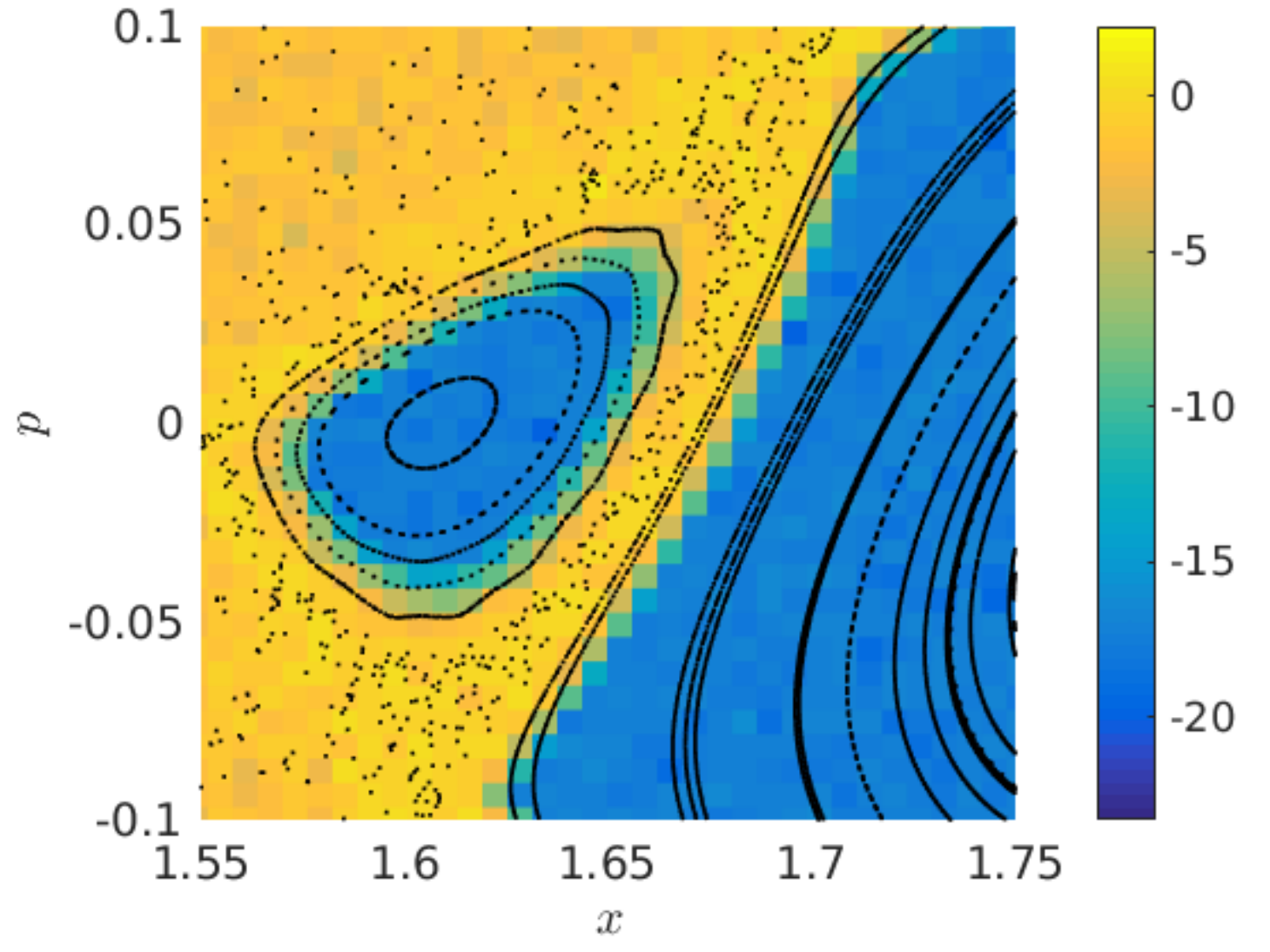}{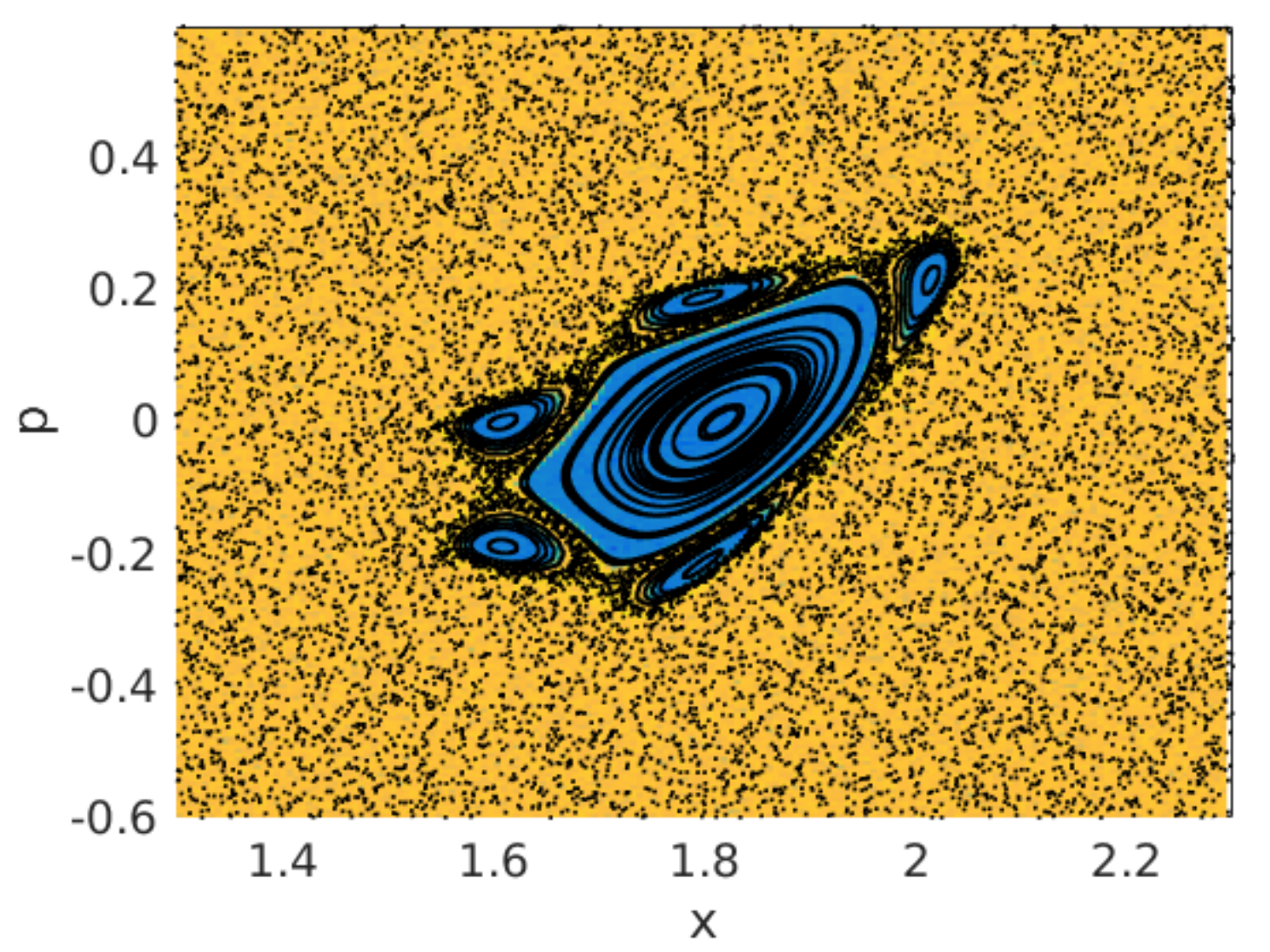}{Color scale plots of the $\log _{10}(\mathcal{H})$ of Husimi distribution on small island for $t=10^4$. For comparison classical trajectories are in black dots in all sub-figures, the outer circle form the island's boundary. (a) $N=2^{17}$ (b) $N=2^{19}$ , and (c) $N=2^{21}$. In (d) we show the entire region of phase space where sticking takes place. The color scale is the same for all sub figures. It is clear that for larger $N$ islands are less occupied by the wave packet (darker color means the Husimi function is smaller). The initial wave-packet was placed near hyperbolic point as in Fig. \ref{fig:fig1}}{island}{0.4}.

\InsertFigTwo{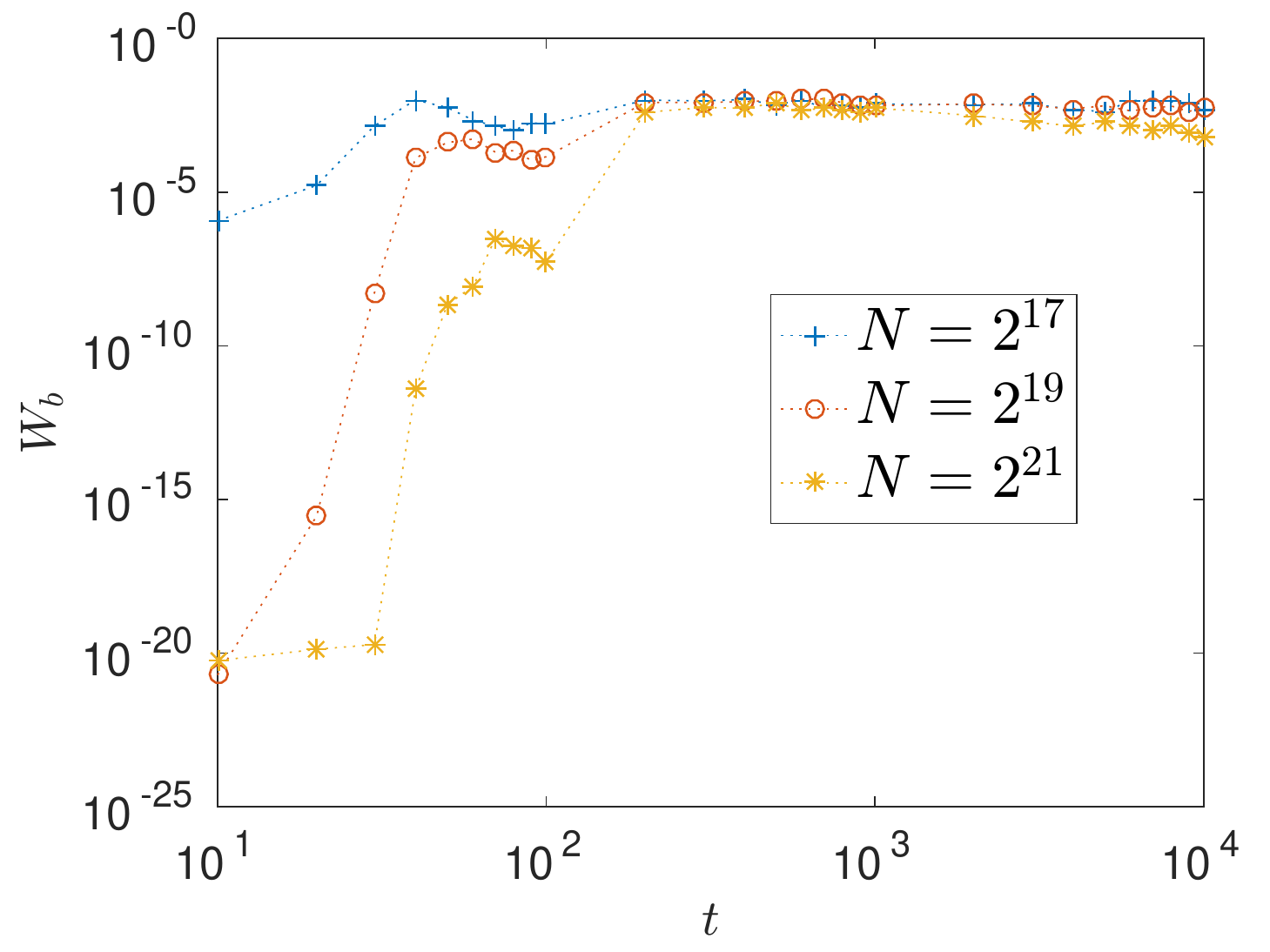}{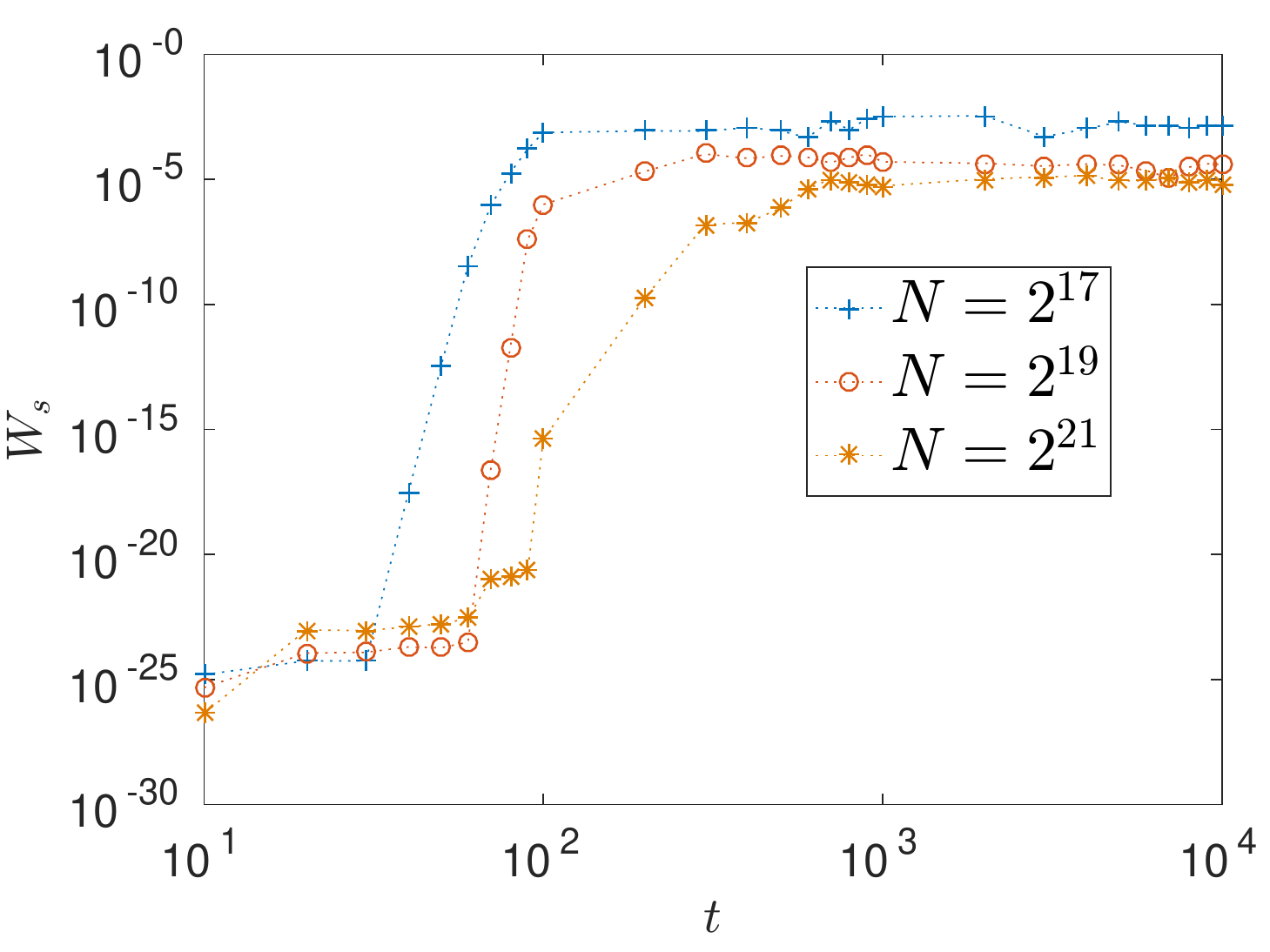}{(a) The weight inside the big island $W_b$ for several values of $N$. (b) The weight inside the small island $W_s$. 
}{weightsHyp}{0.4}

\subsection{The region where the sticking takes place}
The sticking takes place in a region around the island. In \Fig{Sticking}(a) this is demonstrated. The selected region is confined by the unstable and stable manifold of the period-1 hyperbolic fixed point all the way to their first homoclinic intersection \cite{Meiss1997}. The weight $W$ is calculated using \Eq{Weq} where the domain $\mathcal{D}$ is the strip presented in \Fig{Sticking} (a), and is presented  in \Fig{Sticking} (b). 
The decay is obvious; note the visible steps for large $t$.
This decay is described in detail by the modified Markov tree presented in Sec. \ref{sec:Markov}.
\InsertFigTwo{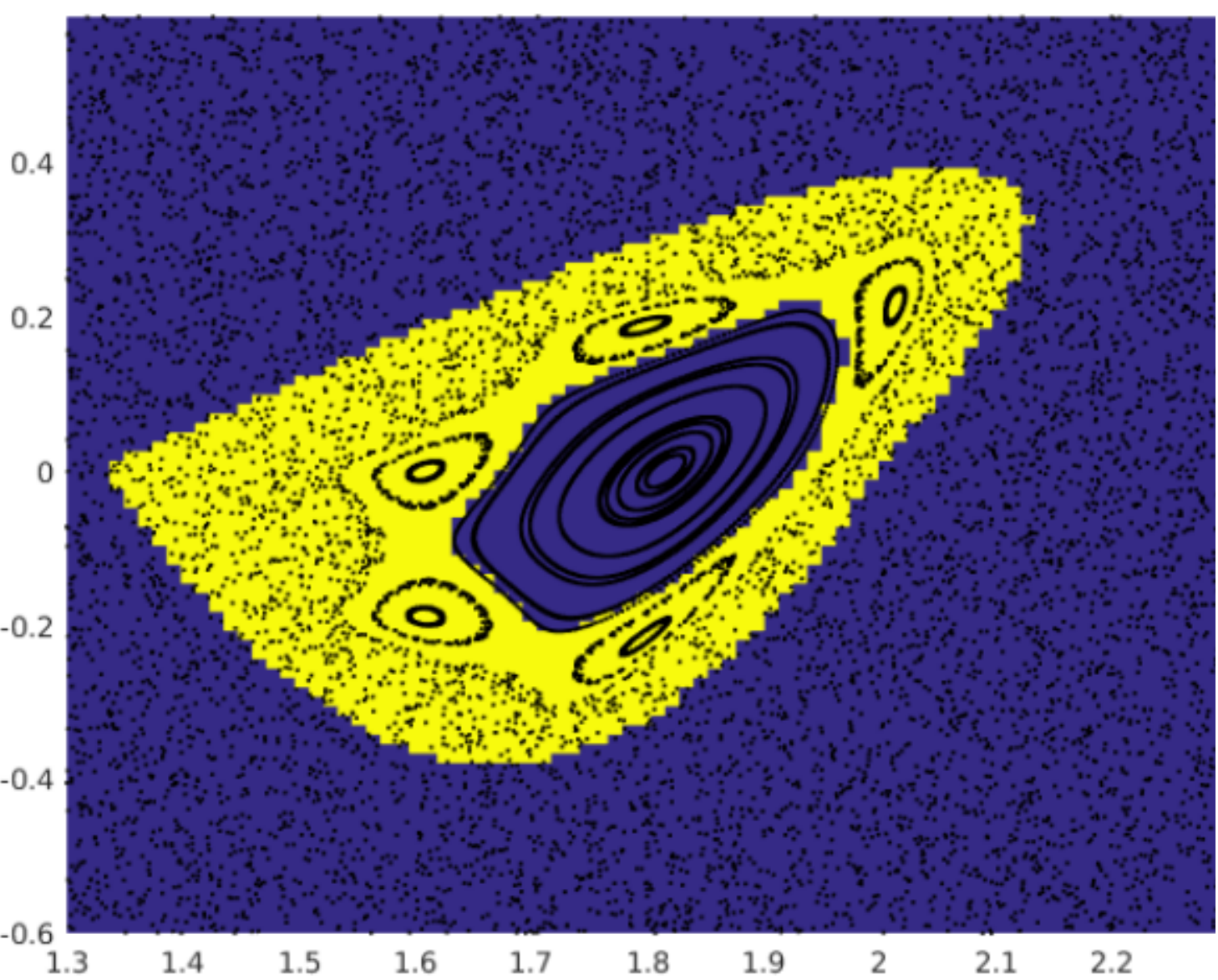}{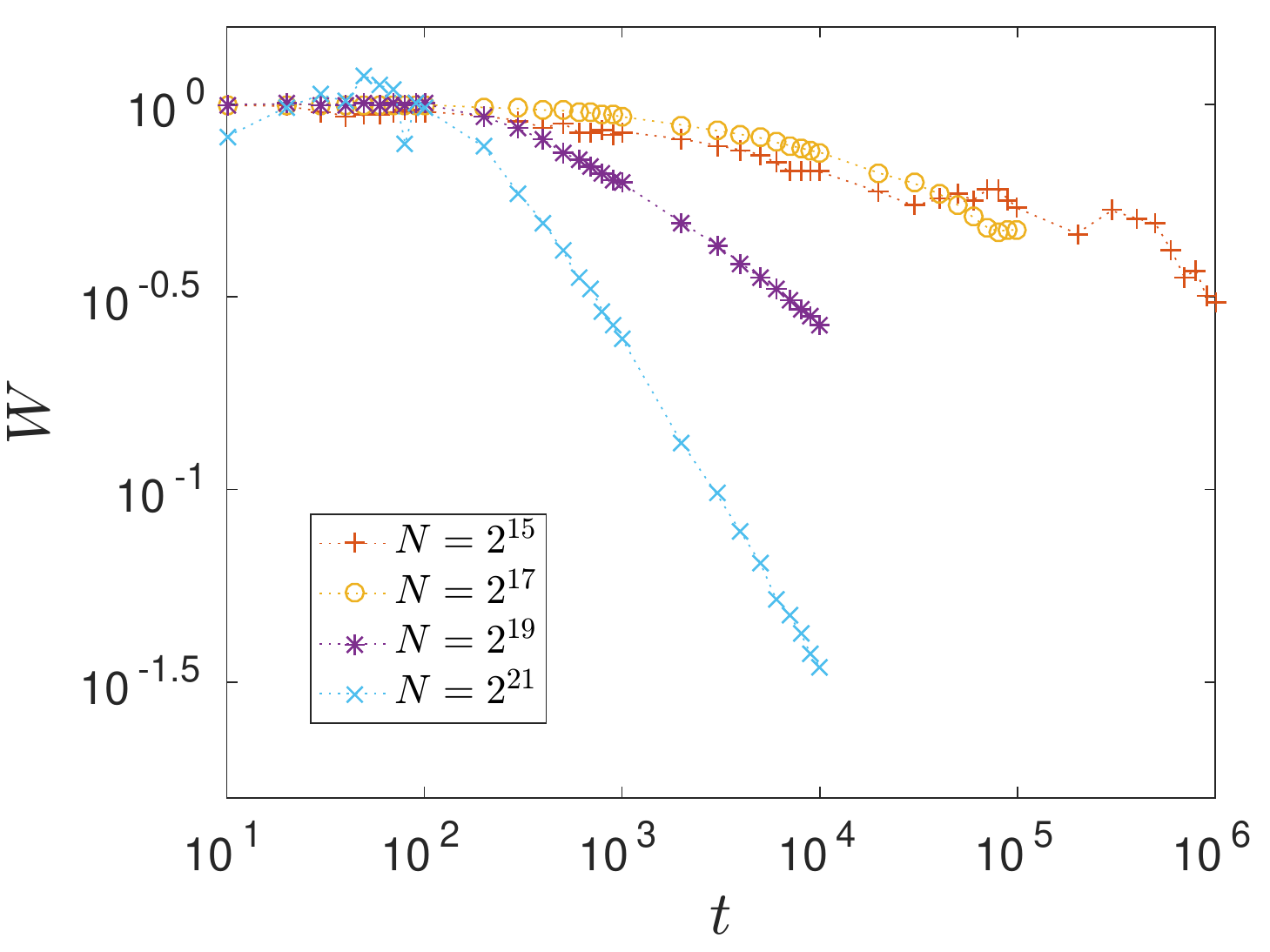}{(a) The sticking strip around the big island. (b) The weight $W$ of this strip for a range of values of $N$.
}{Sticking}{0.4}
 \subsection{Scars} The scars \cite{heller1984bound} in eigenfunctions result in asymptotic trapping, therefore these cannot serve as a mechanism for quantum decay. For this reason we turn to modification of the Markov model.
\subsection{Modification of the Markov model}\label{sec:Markov}
We now try to explain the slowdown qualitatively by using a modification of the Markov model introduced by Meiss and Ott \cite{Meiss1985} and studied in \cite{Alus2017} for the classical case. The transition rates on the tree are defined by $p_{S\to S'}=\frac{\Delta W_{S,S'}}{A_S}$. Where $A_S$ is the accessible region and $W_{S,S'}$ is the flux through a turnstile area between two neighboring regions.
In \cite{michler2012}  the quantum asymptotic transition weight, that is the projection of the wave function started on one side of the turnstile on a region located on the other side after some time, and then averaged over time, was computed. It turns out to agree with a simple random transition model. It was found that this weight should be changed with $\hbar$ via the function 
\begin{equation}\label{eq:RMT}
\frac{\frac{\Delta W}{\hbar}}{1+\frac{\Delta W}{\hbar}}. 
\end{equation}
Inspired by their work we conjecture that for each turnstile the corresponding function is
\begin{equation}\label{eq:scaling}
f(\frac{\Delta W}{\hbar})=\tanh(\frac{\Delta W}{\hbar}).
\end{equation}
Both \Eq{RMT} and \Eq{scaling} converge to the correct limits in the extreme values of $\hbar$. We checked that both \Eq{RMT} and \Eq{scaling} give qualitatively similar results. Using \Eq{scaling} changes the rates such that
\begin{equation}\label{eq:rates}
p_{S\to S'}=\frac{\Delta W_{S,S'}}{A_S}\tanh(\frac{\Delta W}{\hbar}).
\end{equation}

We now build a Markov matrix in a manner similar to what described in \cite{Alus2017}. Here we assume that the Markov property, namely that each of the turnstiles acts independently, holds also quantum mechanically, and here it provides a transition probability (rather an amplitude). This is a result of the fact that the turnstiles are separated by chaotic regions. The rate $p_{1\to \emptyset'}$ is set arbitrarily to be $0.1$, and $A_1$  is set to be the size of the major island discussed above. This determines $\Delta W_{1,\emptyset}$ via \Eq{rates}. Next all other values found using scalings drawn from distributions of the flux and areas as in \cite{Alus2017}. In that way a matrix describing the master equation can be calculated. Next, the matrix is used to propagate an initial distribution equivalent to detailed balance if $p_{1\to \emptyset'}=0$ \cite{Cristadoro2008}, and
\begin{equation}
 \rho_S(0) p_{S\to DS}= \rho_{DS}(0) p_{DS\to s} 
\end{equation}
 This is done for several values of $\hbar$ for the same realization of transition rates. The result is presented in Fig. \ref{fig:Tree}. It shows that small fluctuations in the $\hbar\to 0$ limit turn into regions of exponential decay when $\hbar$ is increased. This happens since when $\frac{\Delta W}{\hbar}<<1$, then $\tanh(\frac{\Delta W}{\hbar})\sim \frac{\Delta W}{\hbar}$. Since for most entries of the matrix $\frac{\Delta W}{\hbar}<<1$  multiplying $\hbar$ by 2 is the same as dividing almost the entire matrix $\mathcal{W}$ given by \cite{Alus2017} :
\begin{equation}\label{eq:Matrix}
\frac{d\vec{\rho}}{dt}=\mathcal{W}\vec{\rho}\,, \quad 
\mathcal{W}_{S,S'} = p_{S'\to S}-\delta_{S,S'}\sum_{S''}p_{S\to S''},
\end{equation}
 by the same factor. This will cause the eigenvalues to separate, and due to separation of time scales, exponential decay appear for increasing times intervals. In intermediate cases of $\frac{\Delta W}{\hbar}$  the power law found for the classical case turns out to be replaced  by a step structure, presented in \Fig{Tree}.
  The classical power law decay is a result of contribution of many small decay exponents related to high generations of the tree, and it goes forever for infinite tree \cite{Alus2017}. If the tree is truncated at a finite generation, at the time corresponding to that generation, the power law decay stops and exponential decay is found. Finite $\hbar$ eventually leads to such a truncation. The manifestation of the step structure for longer times was found by explicit  calculations.
  It  was found that the dependence of the survival probability on time depends on the choice of $A_1$ for the short time part. The step structure of the survival probability is a result of the modification in the Markov tree and it results of the quantum mechanics. We check explicitly that such structure is also found for a typical matrix generated at random for the Markov tree multiplied by factors that control the step structure. A similar step structure was found for the system we study (see \Fig{Sticking}), e.g. for $N=2^{15}$.
\InsertFig{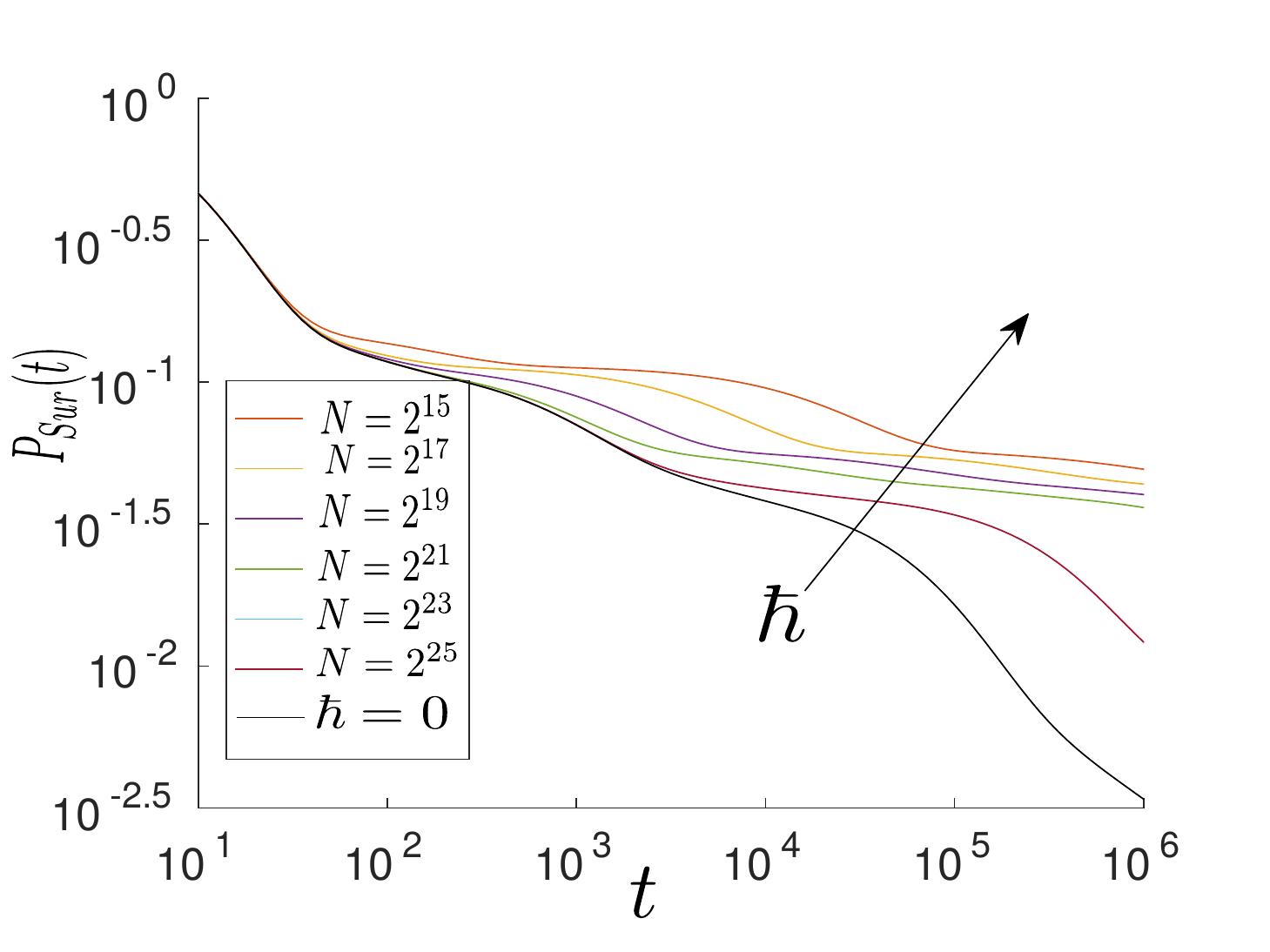}{Survival probabilities over the Markov tree model for several values of $\hbar$. The black curve for $\tanh(\frac{\Delta W}{\hbar})\equiv 1$. The arrow shows the direction of increasing $\hbar$}{Tree}{0.8}

It is important to note that since we are interested in small values of $\hbar$ we keep many generations of the tree where it is pruned and not yet truncated.
\section{Conclusions}\label{sec:Conc}

We have studied both the classical and quantum dynamics for a simple kicked system (the standard map) that classically has mixed phase space. For initial conditions in a portion of the chaotic region that is close enough to the regular region, the phenomenon of sticking leads to a power-law decay with time of the classical correlation function of a simple observable. Quantum mechanically, we observed the same behavior, but with a smaller exponent. We considered various possible explanations of this phenomenon, and settled on a modification of the Meiss--Ott Markov tree model that takes into account quantum limitations on the flux through a turnstile region between regions corresponding to the states of the tree. A natural question is why the correlations decay in spite of the fact that the spectrum is discrete. One should notice that the discreteness of the spectrum is important for time very much longer than the Heisenberg time $\tau_H$ \cite{Berry1991} In our case  $\tau_H \approx (2\pi)^2\cdot 2^{21}\sim 10^6$, while the correlation function is studied for $t<10^4$. Further work is needed to better understand the quantum behavior. Our results can be considered as a first step in such a direction. 
			
{\bf Acknowledgments}: We would like to thank Oded Agam, Arnd B\"{a}cker, Roland Ketzmerick, and James Meiss for fruitful discussions. OA and SF would like to acknowledge partial support of Israel Science Foundation (ISF) grant 931/16. SF thanks the Kavli Inst.~for Theor.~Phys. for its hospitality, where this research was supported in part by the National Science Foundation under Grant  PHY11-25915. MS acknowledges partial support from NSF Grant PHY13-16748.
			
\bibliographystyle{apsrev4-1}
\bibliography{Thermalization}

\begin{thebibliography}{30}%
\makeatletter
\providecommand \@ifxundefined [1]{%
 \@ifx{#1\undefined}
}%
\providecommand \@ifnum [1]{%
 \ifnum #1\expandafter \@firstoftwo
 \else \expandafter \@secondoftwo
 \fi
}%
\providecommand \@ifx [1]{%
 \ifx #1\expandafter \@firstoftwo
 \else \expandafter \@secondoftwo
 \fi
}%
\providecommand \natexlab [1]{#1}%
\providecommand \enquote  [1]{``#1''}%
\providecommand \bibnamefont  [1]{#1}%
\providecommand \bibfnamefont [1]{#1}%
\providecommand \citenamefont [1]{#1}%
\providecommand \href@noop [0]{\@secondoftwo}%
\providecommand \href [0]{\begingroup \@sanitize@url \@href}%
\providecommand \@href[1]{\@@startlink{#1}\@@href}%
\providecommand \@@href[1]{\endgroup#1\@@endlink}%
\providecommand \@sanitize@url [0]{\catcode `\\12\catcode `\$12\catcode
  `\&12\catcode `\#12\catcode `\^12\catcode `\_12\catcode `\%12\relax}%
\providecommand \@@startlink[1]{}%
\providecommand \@@endlink[0]{}%
\providecommand \url  [0]{\begingroup\@sanitize@url \@url }%
\providecommand \@url [1]{\endgroup\@href {#1}{\urlprefix }}%
\providecommand \urlprefix  [0]{URL }%
\providecommand \Eprint [0]{\href }%
\providecommand \doibase [0]{http://dx.doi.org/}%
\providecommand \selectlanguage [0]{\@gobble}%
\providecommand \bibinfo  [0]{\@secondoftwo}%
\providecommand \bibfield  [0]{\@secondoftwo}%
\providecommand \translation [1]{[#1]}%
\providecommand \BibitemOpen [0]{}%
\providecommand \bibitemStop [0]{}%
\providecommand \bibitemNoStop [0]{.\EOS\space}%
\providecommand \EOS [0]{\spacefactor3000\relax}%
\providecommand \BibitemShut  [1]{\csname bibitem#1\endcsname}%
\let\auto@bib@innerbib\@empty
\bibitem [{\citenamefont {Lichtenberg}\ and\ \citenamefont
  {Lieberman}(1983)}]{lichtenberg1983}%
  \BibitemOpen
  \bibfield  {author} {\bibinfo {author} {\bibfnamefont {A.~J.}\ \bibnamefont
  {Lichtenberg}}\ and\ \bibinfo {author} {\bibfnamefont {M.~A.}\ \bibnamefont
  {Lieberman}},\ }\href@noop {} {\bibfield  {journal} {\bibinfo  {journal}
  {Applied Mathematical Sciences}\ } (\bibinfo {year} {1983})}\BibitemShut
  {NoStop}%
\bibitem [{\citenamefont {Tabor}(1989)}]{Tabor1989}%
  \BibitemOpen
  \bibfield  {author} {\bibinfo {author} {\bibfnamefont {M.}~\bibnamefont
  {Tabor}},\ }\href@noop {} {\emph {\bibinfo {title} {Chaos and integrability
  in nonlinear dynamics: an introduction}}}\ (\bibinfo  {publisher} {Wiley},\
  \bibinfo {year} {1989})\BibitemShut {NoStop}%
\bibitem [{\citenamefont {Zaslavsky}(2005)}]{zaslavsky2005hamiltonian}%
  \BibitemOpen
  \bibfield  {author} {\bibinfo {author} {\bibfnamefont {G.~M.}\ \bibnamefont
  {Zaslavsky}},\ }\href@noop {} {\emph {\bibinfo {title} {Hamiltonian chaos and
  fractional dynamics}}}\ (\bibinfo  {publisher} {Oxford University Press on
  Demand},\ \bibinfo {year} {2005})\BibitemShut {NoStop}%
\bibitem [{\citenamefont {H\'enon}(1969)}]{Henon1969}%
  \BibitemOpen
  \bibfield  {author} {\bibinfo {author} {\bibfnamefont {M.}~\bibnamefont
  {H\'enon}},\ }\href@noop {} {\bibfield  {journal} {\bibinfo  {journal} {Q. J.
  Appl. Math.}\ }\textbf {\bibinfo {volume} {27}},\ \bibinfo {pages} {291}
  (\bibinfo {year} {1969})}\BibitemShut {NoStop}%
\bibitem [{\citenamefont {Zumofen}\ and\ \citenamefont
  {Klafter}(1994)}]{Zumofen1994}%
  \BibitemOpen
  \bibfield  {author} {\bibinfo {author} {\bibfnamefont {G.}~\bibnamefont
  {Zumofen}}\ and\ \bibinfo {author} {\bibfnamefont {J.}~\bibnamefont
  {Klafter}},\ }\href@noop {} {\bibfield  {journal} {\bibinfo  {journal}
  {Europhys. Lett.)}\ }\textbf {\bibinfo {volume} {25}},\ \bibinfo {pages}
  {565} (\bibinfo {year} {1994})}\BibitemShut {NoStop}%
\bibitem [{\citenamefont {Chirikov}(1979)}]{Chirikov1979a}%
  \BibitemOpen
  \bibfield  {author} {\bibinfo {author} {\bibfnamefont {B.}~\bibnamefont
  {Chirikov}},\ }\href {https://doi.org/10.1016/0370-1573(79)90023-1}
  {\bibfield  {journal} {\bibinfo  {journal} {Physics Reports}\ }\textbf
  {\bibinfo {volume} {52}},\ \bibinfo {pages} {263} (\bibinfo {year}
  {1979})}\BibitemShut {NoStop}%
\bibitem [{\citenamefont {Katz-Saporta}\ and\ \citenamefont
  {Efrati}(2017)}]{Katz2017}%
  \BibitemOpen
  \bibfield  {author} {\bibinfo {author} {\bibfnamefont {O.}~\bibnamefont
  {Katz-Saporta}}\ and\ \bibinfo {author} {\bibfnamefont {E.}~\bibnamefont
  {Efrati}},\ }\href@noop {} {\bibfield  {journal} {\bibinfo  {journal} {arXiv
  preprint arXiv:1706.09868}\ } (\bibinfo {year} {2017})}\BibitemShut {NoStop}%
\bibitem [{\citenamefont {Channon}\ and\ \citenamefont
  {Lebowitz}(1980)}]{Channon1980}%
  \BibitemOpen
  \bibfield  {author} {\bibinfo {author} {\bibfnamefont {S.~R.}\ \bibnamefont
  {Channon}}\ and\ \bibinfo {author} {\bibfnamefont {J.~L.}\ \bibnamefont
  {Lebowitz}},\ }\href {\doibase 10.1111/j.1749-6632.1980.tb29680.x} {\bibfield
   {journal} {\bibinfo  {journal} {Annals of the New York Academy of Sciences}\
  }\textbf {\bibinfo {volume} {357}},\ \bibinfo {pages} {108} (\bibinfo {year}
  {1980})}\BibitemShut {NoStop}%
\bibitem [{\citenamefont {Chirikov}\ and\ \citenamefont
  {Shepelyansky}(1984)}]{Chirikov1984}%
  \BibitemOpen
  \bibfield  {author} {\bibinfo {author} {\bibfnamefont {B.}~\bibnamefont
  {Chirikov}}\ and\ \bibinfo {author} {\bibfnamefont {D.}~\bibnamefont
  {Shepelyansky}},\ }\href
  {http://www.sciencedirect.com/science/article/pii/0167278984901404}
  {\bibfield  {journal} {\bibinfo  {journal} {Physica D}\ }\textbf {\bibinfo
  {volume} {13}},\ \bibinfo {pages} {395} (\bibinfo {year} {1984})}\BibitemShut
  {NoStop}%
\bibitem [{\citenamefont {Meiss}\ and\ \citenamefont {Ott}(1985)}]{Meiss1985}%
  \BibitemOpen
  \bibfield  {author} {\bibinfo {author} {\bibfnamefont {J.}~\bibnamefont
  {Meiss}}\ and\ \bibinfo {author} {\bibfnamefont {E.}~\bibnamefont {Ott}},\
  }\href {http://dx.doi.org/10.1103/PhysRevLett.55.2741} {\bibfield  {journal}
  {\bibinfo  {journal} {Phys. Rev. Lett.}\ }\textbf {\bibinfo {volume} {55}},\
  \bibinfo {pages} {2741} (\bibinfo {year} {1985})}\BibitemShut {NoStop}%
\bibitem [{\citenamefont {Karney}\ \emph {et~al.}(1982)\citenamefont {Karney},
  \citenamefont {Rechester},\ and\ \citenamefont {White}}]{Karney1982}%
  \BibitemOpen
  \bibfield  {author} {\bibinfo {author} {\bibfnamefont {C.}~\bibnamefont
  {Karney}}, \bibinfo {author} {\bibfnamefont {A.}~\bibnamefont {Rechester}}, \
  and\ \bibinfo {author} {\bibfnamefont {R.}~\bibnamefont {White}},\ }\href
  {http://dx.doi.org/10.1016/0167-2789(82)90045-8} {\bibfield  {journal}
  {\bibinfo  {journal} {Physica D: Nonlinear Phenomena}\ }\textbf {\bibinfo
  {volume} {4}},\ \bibinfo {pages} {425 } (\bibinfo {year} {1982})}\BibitemShut
  {NoStop}%
\bibitem [{\citenamefont {Karney}(1983)}]{Karney1983}%
  \BibitemOpen
  \bibfield  {author} {\bibinfo {author} {\bibfnamefont {C.}~\bibnamefont
  {Karney}},\ }\href {https://doi.org/10.1016/0167-2789(83)90232-4} {\bibfield
  {journal} {\bibinfo  {journal} {Physica D}\ }\textbf {\bibinfo {volume}
  {8}},\ \bibinfo {pages} {360} (\bibinfo {year} {1983})}\BibitemShut {NoStop}%
\bibitem [{\citenamefont {MacKay}\ \emph {et~al.}(1984)\citenamefont {MacKay},
  \citenamefont {Meiss},\ and\ \citenamefont {Percival}}]{MacKay1984}%
  \BibitemOpen
  \bibfield  {author} {\bibinfo {author} {\bibfnamefont {R.}~\bibnamefont
  {MacKay}}, \bibinfo {author} {\bibfnamefont {J.}~\bibnamefont {Meiss}}, \
  and\ \bibinfo {author} {\bibfnamefont {I.}~\bibnamefont {Percival}},\ }\href
  {http://dx.doi.org/10.1016/0167-2789(84)90270-7} {\bibfield  {journal}
  {\bibinfo  {journal} {Physica D}\ }\textbf {\bibinfo {volume} {13}},\
  \bibinfo {pages} {55} (\bibinfo {year} {1984})}\BibitemShut {NoStop}%
\bibitem [{\citenamefont {Zaslavsky}\ \emph {et~al.}(1997)\citenamefont
  {Zaslavsky}, \citenamefont {Edelman},\ and\ \citenamefont
  {Niyazov}}]{Zaslavsky1997}%
  \BibitemOpen
  \bibfield  {author} {\bibinfo {author} {\bibfnamefont {G.}~\bibnamefont
  {Zaslavsky}}, \bibinfo {author} {\bibfnamefont {M.}~\bibnamefont {Edelman}},
  \ and\ \bibinfo {author} {\bibfnamefont {B.}~\bibnamefont {Niyazov}},\ }\href
  {http://dx.doi.org/10.1063/1.166252} {\bibfield  {journal} {\bibinfo
  {journal} {Chaos}\ }\textbf {\bibinfo {volume} {7}},\ \bibinfo {pages} {159}
  (\bibinfo {year} {1997})}\BibitemShut {NoStop}%
\bibitem [{\citenamefont {Ceder}\ and\ \citenamefont {Agam}(2013)}]{Ceder2013}%
  \BibitemOpen
  \bibfield  {author} {\bibinfo {author} {\bibfnamefont {R.}~\bibnamefont
  {Ceder}}\ and\ \bibinfo {author} {\bibfnamefont {O.}~\bibnamefont {Agam}},\
  }\href {https://doi.org/10.1103/PhysRevE.87.012918} {\bibfield  {journal}
  {\bibinfo  {journal} {Phys. Rev. E}\ }\textbf {\bibinfo {volume} {87}},\
  \bibinfo {pages} {012918} (\bibinfo {year} {2013})}\BibitemShut {NoStop}%
\bibitem [{\citenamefont {Ishizaki}\ \emph {et~al.}(1991)\citenamefont
  {Ishizaki}, \citenamefont {Horita}, \citenamefont {Kobayashi},\ and\
  \citenamefont {Mori}}]{Ishizaki1991}%
  \BibitemOpen
  \bibfield  {author} {\bibinfo {author} {\bibfnamefont {R.}~\bibnamefont
  {Ishizaki}}, \bibinfo {author} {\bibfnamefont {T.}~\bibnamefont {Horita}},
  \bibinfo {author} {\bibfnamefont {T.}~\bibnamefont {Kobayashi}}, \ and\
  \bibinfo {author} {\bibfnamefont {H.}~\bibnamefont {Mori}},\ }\href@noop {}
  {\bibfield  {journal} {\bibinfo  {journal} {Prog. Theor. Phys.}\ }\textbf
  {\bibinfo {volume} {85}},\ \bibinfo {pages} {1013} (\bibinfo {year}
  {1991})}\BibitemShut {NoStop}%
\bibitem [{\citenamefont {Manos}\ and\ \citenamefont
  {Robnik}(2014)}]{Manos2014}%
  \BibitemOpen
  \bibfield  {author} {\bibinfo {author} {\bibfnamefont {T.}~\bibnamefont
  {Manos}}\ and\ \bibinfo {author} {\bibfnamefont {M.}~\bibnamefont {Robnik}},\
  }\href {\doibase 10.1103/PhysRevE.89.022905} {\bibfield  {journal} {\bibinfo
  {journal} {Phys. Rev. E}\ }\textbf {\bibinfo {volume} {89}},\ \bibinfo
  {pages} {022905} (\bibinfo {year} {2014})}\BibitemShut {NoStop}%
\bibitem [{\citenamefont {Alus}\ \emph {et~al.}(2017)\citenamefont {Alus},
  \citenamefont {Fishman},\ and\ \citenamefont {Meiss}}]{Alus2017}%
  \BibitemOpen
  \bibfield  {author} {\bibinfo {author} {\bibfnamefont {O.}~\bibnamefont
  {Alus}}, \bibinfo {author} {\bibfnamefont {S.}~\bibnamefont {Fishman}}, \
  and\ \bibinfo {author} {\bibfnamefont {J.~D.}\ \bibnamefont {Meiss}},\
  }\href@noop {} {\bibfield  {journal} {\bibinfo  {journal} {Physical Review
  E}\ }\textbf {\bibinfo {volume} {96}},\ \bibinfo {pages} {032204} (\bibinfo
  {year} {2017})}\BibitemShut {NoStop}%
\bibitem [{\citenamefont {Alus}\ \emph {et~al.}(2014)\citenamefont {Alus},
  \citenamefont {Fishman},\ and\ \citenamefont {Meiss}}]{Alus2014}%
  \BibitemOpen
  \bibfield  {author} {\bibinfo {author} {\bibfnamefont {O.}~\bibnamefont
  {Alus}}, \bibinfo {author} {\bibfnamefont {S.}~\bibnamefont {Fishman}}, \
  and\ \bibinfo {author} {\bibfnamefont {J.}~\bibnamefont {Meiss}},\ }\href
  {\doibase 10.1103/PhysRevE.90.062923} {\bibfield  {journal} {\bibinfo
  {journal} {Phys. Rev. E}\ }\textbf {\bibinfo {volume} {90}},\ \bibinfo
  {pages} {062923} (\bibinfo {year} {2014})}\BibitemShut {NoStop}%
\bibitem [{\citenamefont {Cristadoro}\ and\ \citenamefont
  {Ketzmerick}(2008)}]{Cristadoro2008}%
  \BibitemOpen
  \bibfield  {author} {\bibinfo {author} {\bibfnamefont {G.}~\bibnamefont
  {Cristadoro}}\ and\ \bibinfo {author} {\bibfnamefont {R.}~\bibnamefont
  {Ketzmerick}},\ }\href
  {http://link.aps.org/doi/10.1103/PhysRevLett.100.184101} {\bibfield
  {journal} {\bibinfo  {journal} {Phys. Rev. Lett.}\ }\textbf {\bibinfo
  {volume} {100}},\ \bibinfo {pages} {184101} (\bibinfo {year}
  {2008})}\BibitemShut {NoStop}%
\bibitem [{\citenamefont {Frahm}\ and\ \citenamefont
  {Shepelyansky}(2013)}]{Frahm2013}%
  \BibitemOpen
  \bibfield  {author} {\bibinfo {author} {\bibfnamefont {K.~M.}\ \bibnamefont
  {Frahm}}\ and\ \bibinfo {author} {\bibfnamefont {D.~L.}\ \bibnamefont
  {Shepelyansky}},\ }\href@noop {} {\bibfield  {journal} {\bibinfo  {journal}
  {The European Physical Journal B}\ }\textbf {\bibinfo {volume} {86}},\
  \bibinfo {pages} {322} (\bibinfo {year} {2013})}\BibitemShut {NoStop}%
\bibitem [{\citenamefont {Balazs}\ and\ \citenamefont
  {Voros}(1986)}]{Balazs1986}%
  \BibitemOpen
  \bibfield  {author} {\bibinfo {author} {\bibfnamefont {N.}~\bibnamefont
  {Balazs}}\ and\ \bibinfo {author} {\bibfnamefont {A.}~\bibnamefont {Voros}},\
  }\href@noop {} {\bibfield  {journal} {\bibinfo  {journal} {Physics reports}\
  }\textbf {\bibinfo {volume} {143}},\ \bibinfo {pages} {109} (\bibinfo {year}
  {1986})}\BibitemShut {NoStop}%
\bibitem [{\citenamefont {Balazs}\ and\ \citenamefont
  {Voros}(1989)}]{Balazs1989}%
  \BibitemOpen
  \bibfield  {author} {\bibinfo {author} {\bibfnamefont {N.~L.}\ \bibnamefont
  {Balazs}}\ and\ \bibinfo {author} {\bibfnamefont {A.}~\bibnamefont {Voros}},\
  }\href@noop {} {\bibfield  {journal} {\bibinfo  {journal} {Annals of
  Physics}\ }\textbf {\bibinfo {volume} {190}},\ \bibinfo {pages} {1} (\bibinfo
  {year} {1989})}\BibitemShut {NoStop}%
\bibitem [{\citenamefont {Hannay}\ and\ \citenamefont
  {Berry}(1980)}]{Hannay1980}%
  \BibitemOpen
  \bibfield  {author} {\bibinfo {author} {\bibfnamefont {J.}~\bibnamefont
  {Hannay}}\ and\ \bibinfo {author} {\bibfnamefont {M.~V.}\ \bibnamefont
  {Berry}},\ }\href@noop {} {\bibfield  {journal} {\bibinfo  {journal} {Physica
  D: Nonlinear Phenomena}\ }\textbf {\bibinfo {volume} {1}},\ \bibinfo {pages}
  {267} (\bibinfo {year} {1980})}\BibitemShut {NoStop}%
\bibitem [{\citenamefont {Meiss}(1997)}]{Meiss1997}%
  \BibitemOpen
  \bibfield  {author} {\bibinfo {author} {\bibfnamefont {J.}~\bibnamefont
  {Meiss}},\ }\href {http://dx.doi.org/10.1063/1.166245} {\bibfield  {journal}
  {\bibinfo  {journal} {Chaos}\ }\textbf {\bibinfo {volume} {7}},\ \bibinfo
  {pages} {139} (\bibinfo {year} {1997})}\BibitemShut {NoStop}%
\bibitem [{\citenamefont {Bittrich}\ \emph {et~al.}(2014)\citenamefont
  {Bittrich}, \citenamefont {B{\"a}cker},\ and\ \citenamefont
  {Ketzmerick}}]{bittrich2014temporal}%
  \BibitemOpen
  \bibfield  {author} {\bibinfo {author} {\bibfnamefont {L.}~\bibnamefont
  {Bittrich}}, \bibinfo {author} {\bibfnamefont {A.}~\bibnamefont
  {B{\"a}cker}}, \ and\ \bibinfo {author} {\bibfnamefont {R.}~\bibnamefont
  {Ketzmerick}},\ }\href@noop {} {\bibfield  {journal} {\bibinfo  {journal}
  {Physical Review E}\ }\textbf {\bibinfo {volume} {89}},\ \bibinfo {pages}
  {032922} (\bibinfo {year} {2014})}\BibitemShut {NoStop}%
\bibitem [{\citenamefont {B{\"a}cker}\ \emph {et~al.}(2005)\citenamefont
  {B{\"a}cker}, \citenamefont {Ketzmerick},\ and\ \citenamefont
  {Monastra}}]{Backer2005}%
  \BibitemOpen
  \bibfield  {author} {\bibinfo {author} {\bibfnamefont {A.}~\bibnamefont
  {B{\"a}cker}}, \bibinfo {author} {\bibfnamefont {R.}~\bibnamefont
  {Ketzmerick}}, \ and\ \bibinfo {author} {\bibfnamefont {A.~G.}\ \bibnamefont
  {Monastra}},\ }\href@noop {} {\bibfield  {journal} {\bibinfo  {journal}
  {Physical review letters}\ }\textbf {\bibinfo {volume} {94}},\ \bibinfo
  {pages} {054102} (\bibinfo {year} {2005})}\BibitemShut {NoStop}%
\bibitem [{\citenamefont {Heller}(1984)}]{heller1984bound}%
  \BibitemOpen
  \bibfield  {author} {\bibinfo {author} {\bibfnamefont {E.~J.}\ \bibnamefont
  {Heller}},\ }\href@noop {} {\bibfield  {journal} {\bibinfo  {journal}
  {Physical Review Letters}\ }\textbf {\bibinfo {volume} {53}},\ \bibinfo
  {pages} {1515} (\bibinfo {year} {1984})}\BibitemShut {NoStop}%
\bibitem [{\citenamefont {Michler}\ \emph {et~al.}(2012)\citenamefont
  {Michler}, \citenamefont {B{\"a}cker}, \citenamefont {Ketzmerick},
  \citenamefont {St{\"o}ckmann},\ and\ \citenamefont {Tomsovic}}]{michler2012}%
  \BibitemOpen
  \bibfield  {author} {\bibinfo {author} {\bibfnamefont {M.}~\bibnamefont
  {Michler}}, \bibinfo {author} {\bibfnamefont {A.}~\bibnamefont {B{\"a}cker}},
  \bibinfo {author} {\bibfnamefont {R.}~\bibnamefont {Ketzmerick}}, \bibinfo
  {author} {\bibfnamefont {H.-J.}\ \bibnamefont {St{\"o}ckmann}}, \ and\
  \bibinfo {author} {\bibfnamefont {S.}~\bibnamefont {Tomsovic}},\ }\href@noop
  {} {\bibfield  {journal} {\bibinfo  {journal} {Physical Review Letters}\
  }\textbf {\bibinfo {volume} {109}},\ \bibinfo {pages} {234101} (\bibinfo
  {year} {2012})}\BibitemShut {NoStop}%
\bibitem [{\citenamefont {Berry}(1991)}]{Berry1991}%
  \BibitemOpen
  \bibfield  {author} {\bibinfo {author} {\bibfnamefont {M.~V.}\ \bibnamefont
  {Berry}},\ }\href@noop {} {\bibfield  {journal} {\bibinfo  {journal} {Les
  Houches Summer School "chaos and quantum physics"}\ } (\bibinfo {year}
  {1991})}\BibitemShut {NoStop}%
\end{thebibliography}%
\end{document}